\documentclass[12pt,preprint]{aastex}
\shortauthors{McGreer, Becker, Helfand, \& White}
\shorttitle{$z>6$ Radio-Loud Quasar}

\newcommand\hyperz{\texttt{hyperz}}

\begin{document}
\title{Discovery of a $z=6.1$ Radio-Loud Quasar in the NDWFS}
\author{
Ian D. McGreer\altaffilmark{1},
Robert~H.~Becker\altaffilmark{2,3},
David J. Helfand\altaffilmark{1},
Richard~L.~White\altaffilmark{4}
}

\altaffiltext{1}{Dept. of Astronomy, Columbia University, New York, NY 10027}
\altaffiltext{2}{Physics Dept., University of California, Davis, CA 95616}
\altaffiltext{3}{IGPP/Lawrence Livermore National Laboratory}
\altaffiltext{4}{Space Telescope Science Institute, Baltimore, MD 21218}

\begin{abstract}
From examination of only 4 deg$^2$ of sky in the NOAO Deep Wide-Field Survey
(NDWFS) region, we have identified the first radio-loud quasar at a redshift 
$z>6$. The object, FIRST J1427385+331241, was discovered by matching 
the FLAMEX IR survey to FIRST survey radio sources with NDWFS 
counterparts. One candidate $z>6$ quasar was found, and spectroscopy with the 
Keck II telescope confirmed its identification, yielding a redshift $z=6.12$. 
The object is a Broad Absorption Line (BAL) quasar with an optical luminosity 
of $M_B\sim-26.9$ and a radio-to-optical flux ratio $\sim 60$. 
Two \ion{Mg}{2} absorptions systems are present at redshifts of $z=2.18$
and $z=2.20$. We briefly discuss the implications of this discovery for the 
high-redshift quasar population.

\end{abstract}

\section{Introduction}

High-redshift quasars provide both interesting constraints on the growth
of the first supermassive black holes, and light sources with which
to probe the ionization history of the Universe. The Sloan Digital Sky Survey 
(SDSS) broke the $z=6$ barrier after covering its first 1550 deg$^2$ \citep{Fan01}
and has subsequently identified a total of nine objects with $z \geq 6$ 
drawn from sky coverage of 6550 deg$^2$ \citep{Fan01,Fan03,Fan04,Fan06}. 
Owing to the relatively bright limiting $z'$-band magnitude of the SDSS survey,
all of these quasars are luminous, with $M_B<-26.5$. \citet{Fan04} derive 
estimates for the space density ($6\pm 2 \times 10^{-10}$ Mpc$^{-3}$ 
for $M_{1450}<-26.7$) and luminosity function using nine SDSS-discovered 
quasars with $z>5.7$, concluding $\Psi (L) \sim L^{-3.2\pm0.7}$. This steep 
luminosity function suggests deeper surveys could locate large numbers of 
high-redshift objects.

All of the $z>6$ quasars discovered to date are radio-quiet, with ratios of
radio to optical flux $<10$. One source has been detected at 20~cm 
with a flux density of $55\mu$Jy, while a second has a measured value of 
$26\pm 12~\mu$Jy \citep{carilli04}. The most distant quasar known with a 
flux density level of $>1$~mJy is SDSS J083643.85+005453.3 at $z=5.82$. 
However, based on semi-analytic models of dark halo formation, 
\citet{hqb04} predict high surface densities of high-redshift radio-loud 
quasars; in particular, they suggest roughly four $z>6$ quasars per square 
degree should be found in the catalog constructed from Faint Images of the 
Radio Sky at Twenty-cm (FIRST -- \citealt{FIRST}). Motivated by this 
suggestion,  we have matched the FIRST catalog to the NOAO Deep Wide-Field 
Survey (NDWFS -- \citealt{ndwfs}) and the recently released FLAMINGOS 
EXtragalactic Survey (FLAMEX -- \citealt{flamex}), yielding a net survey area 
of 4.1 deg$^2$.  The result is the detection of the first $z>6$ radio-loud 
quasar.

In Section 2, we briefly describe the catalogs used for this project. We
then go on to present our search procedure which produced one good candidate
object (\S 3). Section 4 describes the spectroscopic confirmation of this 
quasar, \S 5 its continuum properties from rest UV to IR, and \S 6 its
absorption line systems. In \S 7 we discuss the implication of this discovery 
for the high-redshift quasar population as well as for future searches for 
high-redshift objects. Magnitudes are reported in the Vega system unless
otherwise noted. Photometry in the optical to near-infrared was obtained
from SExtractor \citep{1996A&AS..117..393B} MAG\_AUTO magnitudes. Throughout 
the paper we use the standard cosmological model, with parameters 
$\Omega_{\lambda} = 0.7,\ \Omega_M = 0.3,\ {\rm and}\ H_0 = 70$~km s$^{-1}$ Mpc$^{-1}$ (Spergel et al. 2003).

\section{Survey Data}

\subsection{NDWFS}

Beginning in 1999, the NDWFS \citep{ndwfs} obtained deep ($R \sim 26$) images 
in three optical bands ($B_w$, $R$, and $I$) over two 9 deg$^2$ fields with 
the MOSAIC imager on the Mayall 4-m telescope.  The Survey's Third Data 
Release\footnote{$http://www.noao.edu/noao/noaodeep/DR3/dr3-descr.html$} 
includes calibrated images and source catalogs for the northern (Bo\"otes) 
field.  Separate catalogs were derived for each band using SExtractor; a merged
catalog was also produced which includes the three optical bands along with
partial coverage in the $K$ band.

\subsection{FIRST}

We produced the FIRST survey \citep{FIRST} from data collected between 1993 
and 2004 using the Very Large Array\footnote{
The VLA is a facility of the National Radio Astronomy Observatory which is 
operated by Associated Universities, Inc.  under cooperative agreement with 
the National Science Foundation.} 
in its B configuration operating at 20~cm.
Over 9000 deg$^2$ of the Sloan Digital Sky Survey footprint
was imaged to a sensitivity of 1~mJy, with astrometric accuracy better than
$1^{\prime\prime}$; all images and catalogs are available at the
FIRST website\footnote{$http://sundog.stsci.edu$}. The radio source 
surface density is $\sim 90$ deg$^{-2}$; 906 sources fall within the NDWFS 
Bo\"otes field.

\subsection{FLAMEX}

The FLAMEX Survey \citep{flamex} is the largest survey available in the 
near-IR to $K\sim 20$. It was conducted on the KPNO 2.1-m telescope between 
2001 and 2004, ultimately covering $4.7\ {\rm deg}^2$ of the Bo\"otes region; 
the net overlap with NDWFS is 4.1~deg$^2$. The survey is 50\% complete at 
$K_s = 19.5$; the stated limit at $J$ is $\sim 21.4$. A catalog of 
$\sim 157,000$ $K_s$-selected sources was produced using SExtractor and was 
released through the FLAMEX website\footnote{$http://flamingos.astro.ufl.edu/extragalactic/data.html$}.

\section{Candidate Selection}\label{photoz}

In order to identify high-redshift quasar candidates, we searched the 
NDWFS Bo\"otes survey region for compact FIRST sources, defined as 
having a fitted semi-major axis of less than $2.5^{\prime\prime}$.  There 
are 394 FIRST sources meeting this criterion in the NDWFS field, of which 
168 lie in the intersection of the NDWFS and FLAMEX survey regions.  
Roughly 60\% of the FIRST sources can be identified with NDWFS counterparts; 
this fraction increases to $\sim$ 75\% when FLAMEX is included.

The photometric redshift code \hyperz\ of Bolzonella et al. (2000) was used 
to estimate redshifts for the entire sample of optical/IR radio-source
counterparts. The composite radio-emitting quasar spectral template
from the FIRST Bright Quasar Survey (FBQS, Brotherton et al. 2001) was used
to fit the optical/IR photometry. This template was treated as a non-evolving 
model characteristic of radio-loud and radio-intermediate quasars.  We fit the 
template to the NDWFS and FLAMEX photometry over a range of redshifts 
$0 < z < 10$, and included intrinsic dust absorption as a parameter, using the 
starburst galaxy dust model of Calzetti et al. (2000).  The \hyperz\ code 
also includes Lyman forest absorption according to the model of Madau (1995).

A set of four candidate $z > 4$ radio-loud quasars from the best-fit
\hyperz\ redshift estimates was then examined by eye, using the optical 
images from NDWFS; two of these had putative redshifts greater than six. 
Only the source FIRST J1427385+331241 ($F_{peak}=1.7$mJy) had all of the 
expected properties of a bright $z > 6$ quasar: no emission in the $B_w$ and 
$R$ bands, detection in the $I$ band with a high stellarity ($I$ = 22.1, 
SExtractor CLASS\_STAR = 0.98), and relatively bright infrared fluxes 
($J$ = 18.7, $K_s$ = 17.4).  The offsets between the optical, infrared, and 
radio positions are small ($\la 0.3^{\prime\prime}$) and consistent with 
astrometric uncertainties. The source was thus selected as the prime candidate 
for spectroscopic follow-up observations.

\section{Observations}\label{observations}

\begin{figure}
 \plotone{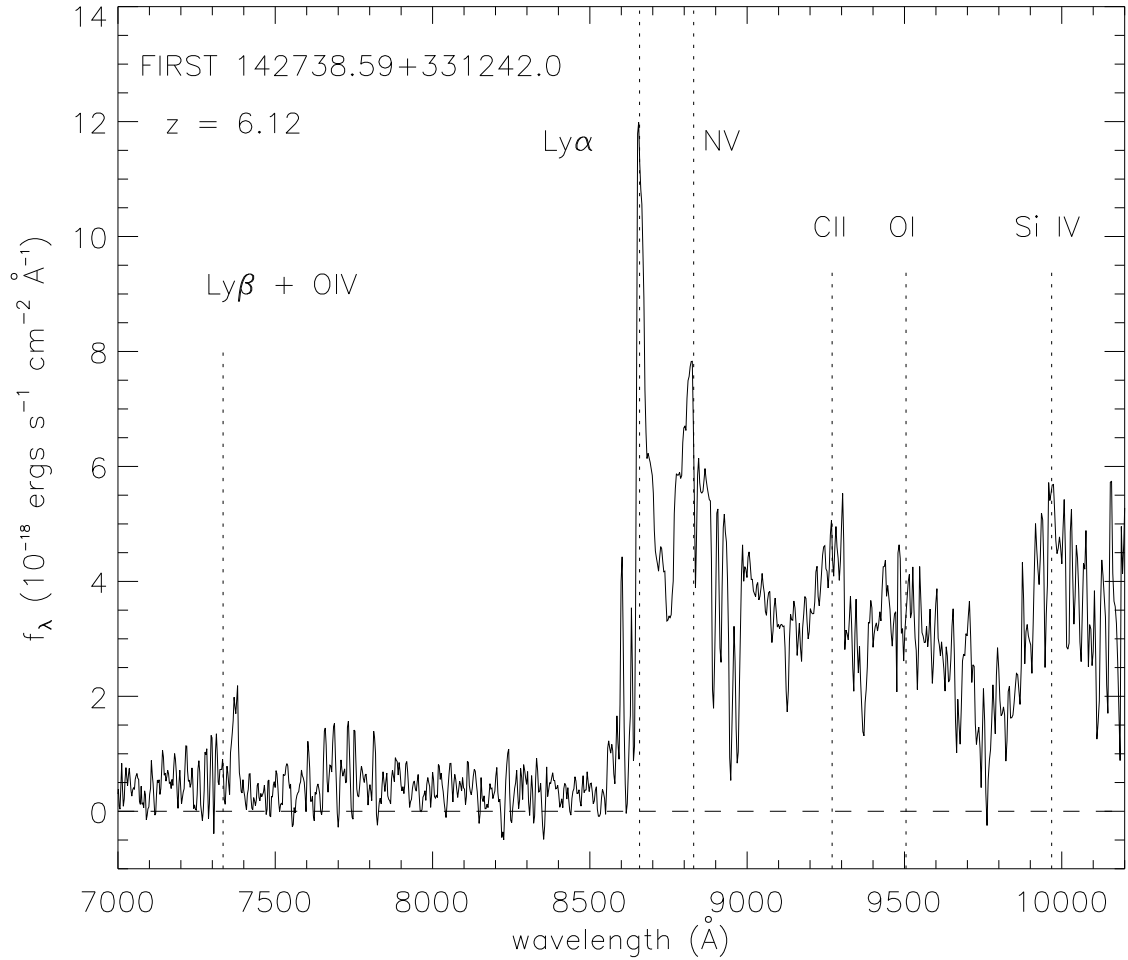}
 \caption
{
 The spectrum of  FIRST J1427385+331241 obtained on 2006 March 5
 using the ESI spectrograph ($R \sim 4000$) on the Keck II telescope. 
 Prominent emission lines are indicated with vertical dashed lines. Strong 
 absorption features blueward of the \ion{N}{5} and \ion{Si}{4} emission 
 lines demonstrate that this is a BAL quasar.
}
 \label{spectrum}
\end{figure}

A discovery spectrum was obtained as the source was rising at dawn on 2006
January 3 at the Keck II telescope using the ESI spectrograph ($R \sim 4000$).
This spectrum was sufficient to identify the source as a $z=6.12$ quasar;
however, the seeing was bad and the source was at a high airmass, so the
S/N was poor.  A higher S/N spectrum was obtained at Keck II on March 5, 
again using the ESI (Figure~\ref{spectrum}). The total exposure time was 
40 minutes under good conditions, with seeing of $0\farcs{8}$.

FIRST J1427385+331241 was observed in $J$ and $K_s$ at the MDM
observatory 2.4-m Hiltner telescope using TIFKAM\footnote{TIFKAM was funded 
by Ohio State University, the MDM consortium, MIT, and NSF grant AST 96-05012. 
NOAO and USNO paid for the development of the ALADDIN arrays and contributed 
the array currently in use in TIFKAM.}
on 2006 April 12 and 15 under non-photometric conditions.  For both bands,
SExtractor was used to detect and measure sources in the $4'\times4'$ 
field.  Calibration was performed using 2MASS stars within the
field.  The quasar was just detected ($S/N \sim 2$) in the $J$ band in a 
20 minute integration,  with a Vega magnitude $J = 19.9 \pm 0.5$. A 28 
minute integration in the $K_s$ band also detected the quasar ($S/N \sim 4$)
with a magnitude $K_s = 17.93 \pm 0.25$. The discrepancy between the MDM
and FLAMEX measurements was resolved by examining the FLAMEX images 
(S.~A.~Stanford, private communication), where it became evident that flux 
from a nearby source ($3\arcsec$ west) was blended with the quasar's flux. The 
standard IRAF DAOPHOT routines were used to construct PSFs from the images 
and to fit simultaneously the quasar and the nearby source. From this process, 
we obtained $J=19.68 \pm 0.05$ and $K_s=17.88 \pm 0.16$ from the FLAMEX 
images, in good agreement with the MDM observations. All photometric
observations of FIRST J1427385+331241 are shown in Table~\ref{photometry}.

The VLA European Large-Area ISO Survey (ELAIS -- \citealt{1999MNRAS.302..222C})
detected this source as ELAISR142738+331242, and reported a 1.4~GHz flux 
density of $1.816 \pm 0.021$ mJy, compared to the $1.73 \pm 0.13$ mJy flux 
density from FIRST. The source is also barely visible in the NRAO VLA Sky Survey
(NVSS -- \citealt{1998AJ....115.1693C}) image of this region, where it is 
merged with a brighter nearby source; the sum of the flux densities of the 
two resolved sources from FIRST is consistent with the reported NVSS flux 
density.  Thus, there is no evidence of large variations at radio wavelengths.

The source is not detected in the Chandra XBootes survey 
(Murray et al. 2005), implying an upper limit in the 0.5-7~keV band
of $4 \times 10^{-15}$~erg cm$^{-2}$ s$^{-1}$. Using the Galactic 
$N_H$ at this location of $1.1 \times 10^{20}$~cm$^{-2}$ and a power law slope
of $\Gamma = 1.9$ yields a rest frame luminosity upper limit
$L(3.5-14~{\rm keV})<1.0 \times 10^{45}$ erg s$^{-1}$, only
a factor of $2-3$ below the X-ray luminosities detected for other high-redshift
(albeit, radio-quiet) quasars (Brandt et al. 2002). The evidence,
presented below, that this object is a BAL could explain its modest 
observed X-ray flux, since intrinsic column densities of up to 
$4 \times 10^{23}$ cm$^{-2}$ have been seen in such objects
(Gallagher et al. 2002); in this case the luminosity upper limit increases 
by $\sim 50\%$.

We obtained Spitzer IRAC archival images of the Bo\"otes field, from a survey
conducted by \citet{2004ApJS..154...48E}.  The source is detected in all four 
IRAC bands. Measured fluxes from analyzing the Post-BCD images with SExtractor 
are $68 \pm 7 \mu$Jy at $3.6\mu$m, $77 \pm 7 \mu$Jy at $4.5\mu$m, 
$93 \pm 8 \mu$Jy at $5.8\mu$m, and $72 \pm 7 \mu$Jy at $8.0\mu$m. 
The Bo\"otes field has also been surveyed at $24\mu$m by 
\citet{2005ApJ...622L.105H}. A catalog is not yet available, but examination 
of archival images reveals a possible detection of the source, based on a 
small fluctuation ($\sim 1.7\sigma$) in the image at the quasar position. We 
adopt a rough upper limit of $0.2$mJy for the $24\mu$m flux of this source.
These mid-infrared fluxes are comparable to measurements of SDSS $z>5.8$ 
quasars from the Spitzer survey conducted by Jiang et al. (2006). From the
$24\mu$m upper limit we see no evidence for a hot dust mass
(see Figure~\ref{sed}) as was found for 11 out of 13 of the quasars observed 
by Jiang et al. (see their Figure 3).  The ratio of rest-frame near-IR to 
optical luminosities for FIRST J1427385+331241 is also low, 
$\log(\nu{\rm L}_\nu[3.5\mu{\rm m}]/\nu{\rm L}_\nu[4400\mbox{\AA}]) \la -0.4$.
Jiang et al. compare the two quasars in their sample showing similarly weak NIR
fluxes to low redshift samples and find that such cases are very rare; they
suggest that dust evolution at high redshift may explain why a significant 
fraction ($\sim 20\%$) of $z \sim 6$ quasars are NIR-weak.

\section{Continuum Properties}

\begin{figure}
 \plotone{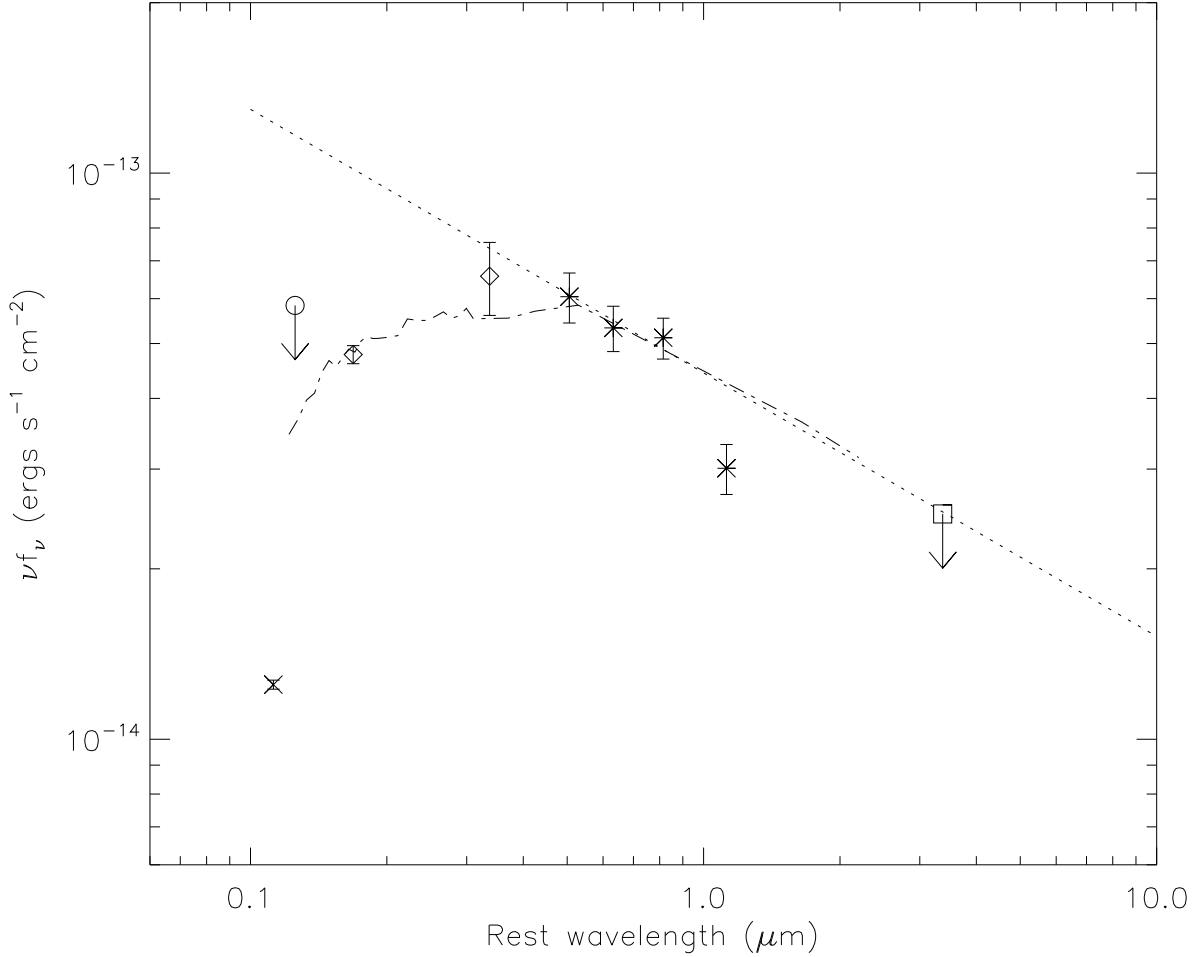}
 \caption
{
 The rest-frame ultraviolet to infrared SED of FIRST J1427385+331241, 
 compiled from multiple surveys. Points with error bars are: 
 NDWFS $I$ (cross), SDSS $z'$ upper limit (circle), FLAMEX $J$,$K_s$ 
 (diamonds), IRAC $3,4,6,8 \mu$m (asterisks), and the MIPS $24\mu$m upper 
 limit (square).  A representative power law with a slope of $\alpha = -0.5$ 
 (dotted line) is fit to the IRAC $3-6\mu$m data and the MIPS $24\mu$m upper 
 limit. The steep drop in flux in the rest-frame ultraviolet suggests that 
 the quasar is reddened by dust extinction. Shown in dot-dashed lines
 is a power law with slope $\alpha=-0.5$ extincted by SMC-like
 dust with $E(B-V) = 0.09$, using the values for SMC dust given in 
 \citet{Prevot} and \citet{Pei92}.
}
 \label{sed}
\end{figure}

In Figure~\ref{sed}, we show the full optical to infrared spectral energy
distribution (SED) of FIRST J1427385+331241. The SED does not resemble
a simple power law, and the large flux ratio between the rest-frame 
ultraviolet and optical range suggests that dust extinction may be
responsible for the observed optical and near-infrared colors ($I-J=2.4$ and
$J-K=1.8$).  We derive the rest-frame luminosity of the quasar by fitting the
redshifted FBQS template to the IRAC measurements, which are the least affected 
by dust. From this process we derive $M_B=-26.9$ and $M_{1450}=-26.4$;
only one known $z>6$ quasar is less luminous: SDSS J1630+4012 has
$M_{1450}=-26.1$ \citep{Fan03}.  Using a template to derive UV/optical 
luminosities ignores dust reddening; the flux levels in our Keck 
spectrum at these wavelengths are much lower, most likely as a consequence of 
slit losses which render these data unreliable for deriving magnitudes. As none 
of the $z\sim6$ SDSS quasars have shown significant dust extinction, using the
template to derive luminosities allows a direct comparison to that sample.

In order to compare the radio and optical properties of 
FIRST J1427385+331241, we adopt the definition of radio loudness from 
Ivezi{\'c} et al. (2002), $R_m=0.4(m-t)$, where $m$ is an optical AB magnitude 
and $t$ is the ``AB radio magnitude'' at 1.4~GHz. This definition does not 
include a $K$-correction, and thus probes different regions of the spectral 
energy distribution at different redshifts. We consider two regions based on 
observations of $z>6$ sources.  At $z\sim6$, the $B$ band is redshifted to the 
near-infrared ($\sim 3{\mu}m$), such that a comparison of the flux at 1.4~GHz 
to the IRAC $3.6\mu$m flux is an approximation to the usual definition of radio 
loudness in terms of the ratio of 5~GHz radio to $B$-band optical flux 
(Kellerman et al. 1989). 
Using an AB magnitude of 19.32 at $3.6\mu$m, the radio loudness for
FIRST J1427385+331241 is $R_{3.6\mu{\rm m}}=1.4$.  We also consider the 
radio-loudness for $z>6$ sources in terms of the rest-frame ultraviolet; 
specifically, to compare with the SDSS quasars, we use the flux at 1450\AA. 
Using the values of $m_{1450}$ obtained from template fitting as described 
above, the radio-loudness of the quasar is $R_{1450} = 1.8$.  The quasar 
SDSS J1148+5251 detected by Carilli et al. (2004) with a flux density at 
1.4~GHz of $55\mu$Jy has $R_{1450} = -0.2$ and $R_{3.6\mu{\rm m}}=-0.4$, so 
FIRST J1427385+331241 is truly the first $z>6$ radio-loud quasar.

FIRST J1427385+331241 is a Broad Absorption Line Quasar (BALQSO), as is 
evident from the strong absorption features blueward of the \ion{N}{5} 
and \ion{Si}{4} emission lines (see Figure~\ref{spectrum}).  BALQSOs are
more intrinsically reddened than non-BALQSOs 
\citep{trump,reichard03b,brotherton01}.  Figure~\ref{sed} 
shows that a simple model using a power law with slope $\alpha=-0.5$
reddened by a Small Magellanic Cloud (SMC, \citealt{Prevot,Pei92}) dust 
model with $E(B-V)=0.1$ gives a good fit to the observed SED.  This dust model 
agrees with the FBQS sample of low-redshift BALQSOs \citep{brotherton01},
where the spectral shape of the low-ionization BALQSO composite was consistent 
with the non-BALQSO composite reddened by SMC dust with $E(B-V)\sim0.1$. 
SDSS J1048+4637, at $z=6.2$, is also a BALQSO; thus the BAL fraction for known 
$z>6$  quasars is $\sim 20\%$, similar to the low-redshift fraction 
($\sim 10-15\%$, \citealt{trump,reichard03b,1991ApJ...373...23W}).

\section{\ion{Mg}{2} absorption}

\begin{figure}
 \plotone{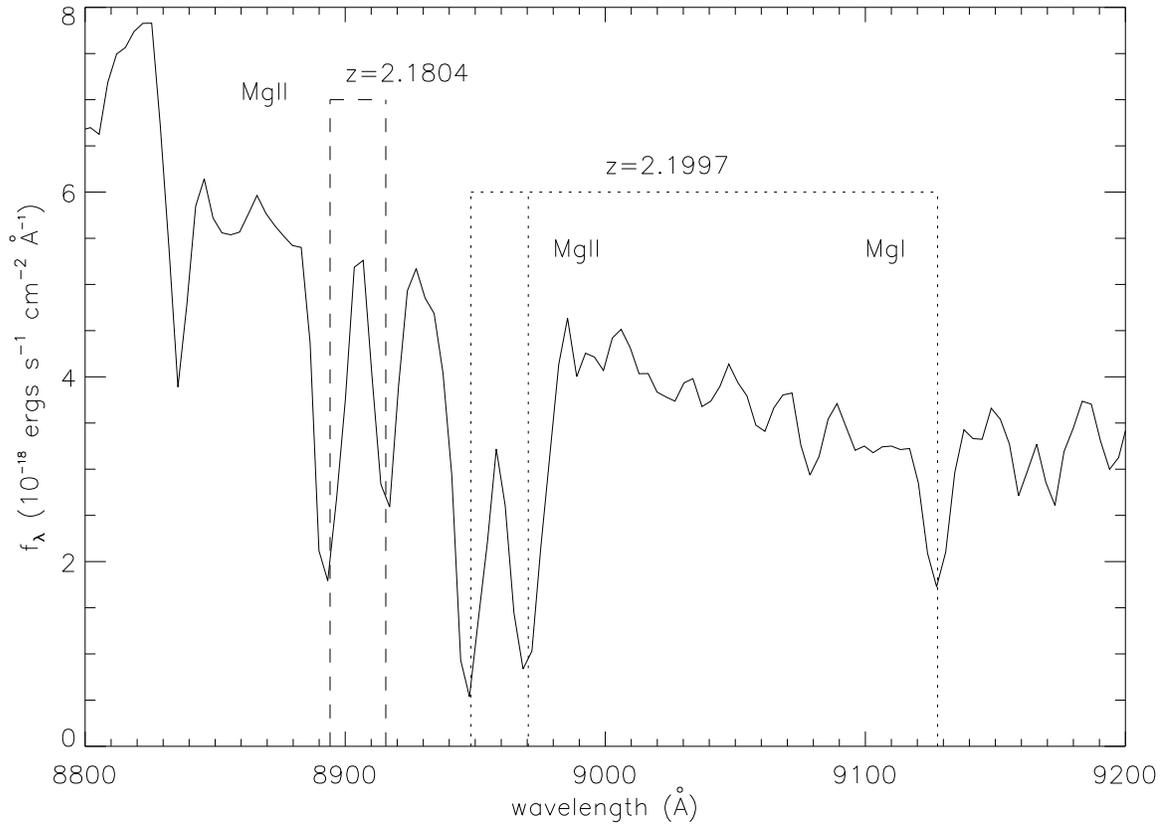}
 \caption
{
 A portion of the Keck spectrum of  FIRST J1427385+331241, revealing the 
 presence of Mg absorption systems at $z_{abs,1}=2.1804 \pm 0.0003$ and 
 $z_{abs,2}=2.1997 \pm 0.0003$.  In both systems, the 
 \ion{Mg}{2} $\lambda\lambda2796, 2803$ doublet is present, while the higher 
 redshift system also shows the weaker \ion{Mg}{1} $\lambda2852$ line.
 }
 \label{mg2}
\end{figure}

The high-S/N Keck spectrum reveals two strong
\ion{Mg}{2} $\lambda\lambda2796, 2803$ absorption systems (Figure~\ref{mg2}).  
The first is at observed wavelengths 8894\AA\ and 8916\AA, giving an 
absorber redshift of $z=2.1804 \pm 0.0003$.  The second shows both the 
\ion{Mg}{2} doublet and the \ion{Mg}{1} $\lambda2852$ line, at observed 
wavelengths 8948\AA, 8970\AA, and 9128\AA\ respectively, yielding an 
absorber at $z=2.1997 \pm 0.0002$.  The velocity separation of these systems 
is $\sim 5800$ km/s; they are unlikely to be part of a single cluster but may
be associated with large-scale structure.

The presence of absorption systems in the line-of-sight raises
the possibility that the quasar is being gravitationally lensed. 
By examining quasars in the 2dF survey, \citet{MenardPeroux} found evidence 
for gravitational magnification of sources with strong 
\ion{Mg}{2}/\ion{Fe}{2} absorbers.  \citet{MurphyLiske} also detected 
magnification of SDSS quasars with intervening DLAs. \citet{MenardPeroux} 
discuss how the absorption systems are responsible for competing effects: 
while the overall brightness of the source is increased by gravitational 
lensing, extinction effects in the absorber's frame will redden the source.
\citet{MurphyLiske} did not find evidence for systematic reddening of
SDSS quasars by DLAs at $z_{abs}\sim3$, but \citet{York06} did find reddening 
as large as $E(B-V)\sim0.1$ in SDSS quasars with \ion{Mg}{2} absorbers
in the interval $1.0<z_{abs}<1.86$. For a single source, especially a BALQSO
which is likely to be intrinsically reddened, it is impossible to tell
whether either of these two effects are present; we simply note that both
the inferred luminosity and the spectral shape of FIRST J1427385+331241 
may be affected by the two \ion{Mg}{2} absorbers.

\section{Discussion}

Approximately 10\% of the normal SDSS quasars are detected in the
FIRST survey (White et al. 2006). The fact that SDSS has found one $z>6$
quasar (all radio quiet) per $\sim 730$ deg$^2$ searched and we 
have found a radio-loud object in a 4 deg$^2$ survey requires
examination,\footnote{One possible explanation is that we are, as previously
documented, just lucky (Becker, Helfand, and White 1992; White, Kinney,
and Becker 1993; Stern et al. 2005).} as a naive estimate would predict our
chances of success to be 4.1~deg$^2/730$~deg$^2 \times 0.1 = 6 \times 10^{-4}$.
Allowing for the possibility that the quasar is being gravitationally
magnified, we might have reached $\sim 1$ mag deeper into the quasar
luminosity function than Fan et al. Even with a steep luminosity function
$\Psi (L) \sim L^{-3.2}$, this would only increase the number density by
a factor of $\sim 20$; the number density of $6\times10^{-10}\ {\rm Mpc}^{-3}$
from \citet{Fan04} would predict $\sim 0.1$ quasars in 4 deg$^2$ in the 
redshift range $6<z<7$, such that, including the radio-loud fraction, our 
chances would have been $\la 1\%$.

\begin{figure}
 \plotone{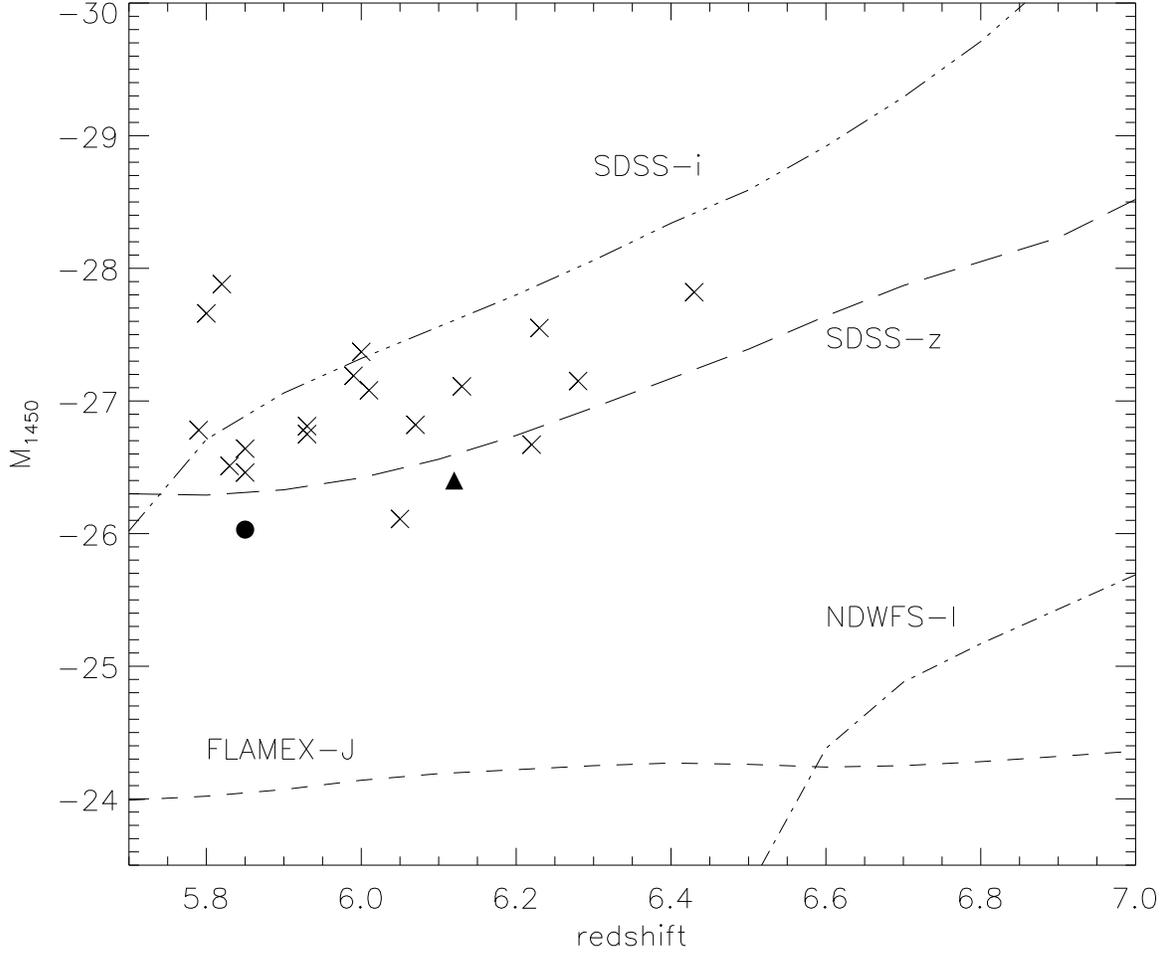}
 \caption 
{
 Detection limits of optical/IR photometric quasar searches obtained by
 redshifting the LBQS optical quasar template \citep{francis91}. The upper two
 lines show the limits of the SDSS $i'$ photometry and the magnitude
 cutoff $z'<20.2$ used for selection of quasar candidates by \citeauthor{Fan01}
 The lower two lines show the faintest luminosities reached by the 
 NDWFS $I$ and FLAMEX $J$ photometry.  Note that the NDWFS $I$ band
 covers nearly the same wavelength range as the SDSS $i'$ and $z'$ bands 
 combined.  Published $z>5.8$ SDSS quasars are shown with crosses. The
 $z=5.85$ quasar found by \citet{cool} using IRAC selection is shown as a 
 filled circle.  FIRST J1427385+331241 is shown as a filled triangle, 
 showing that it is just below the nominal limit of the SDSS quasar search, 
 but well above the NDWFS/FLAMEX detection limits. SDSS J1630+4012 ($z=6.05$) 
 is the faintest $z>6$ quasar ($M_{1450}=-26.1$), yet it was an $8\sigma$ 
 detection, $z'=20.4$ (Fan et al. 2003). The inferred luminosity for 
 FIRST J1427385+331241 is higher ($M_{1450}=-26.4$), but its red
 optical color put it below the SDSS $z'$ detection limit.
}
 \label{maglimits}
\end{figure}

In addition to the predictions of high surface densities of radio-loud
quasars put forth by \citet{hqb04}, we expected that the use of deep optical 
and infrared surveys would allow us to reach less luminous quasars at high 
redshifts.  Figure~\ref{maglimits} shows that the combination of NDWFS and 
FLAMEX should probe much deeper into the luminosity function of high-z quasars 
than the corresponding search conducted using the SDSS. We thus expected to 
find, if anything, a quasar too faint to fall within the SDSS program limits 
and identified by the combination of deep optical/IR imaging with radio 
selection.

This quasar is just below the reach of the SDSS program, as shown in 
Figure~\ref{maglimits}.  The NDWFS $I$ band, which essentially covers the 
combined wavelength range of the SDSS $i'$ and $z'$ bands, easily detected 
this source, as did the FLAMEX $J$. Thus we did not fully utilize the deep 
coverage of the Bo\"otes region surveys in locating this particular source. 
In addition, while radio detection was the primary criterion for our quasar 
search, it was not essential in this case, as the quasar could have been 
identified by its optical and infrared properties alone (as an $R$-band 
dropout); however, the use of radio detection does eliminate contamination from
L and T dwarf stars. This quasar is not particularly underluminous relative to 
other $z>5.8$ quasars (see Figure~\ref{maglimits}); it is the object's red 
optical colors that prevented it from being detected by SDSS. As at 
low-redshift, radio selection has located quasars that are redder than those 
found by optical criteria \citep{white03}.  The improbable discovery of this 
quasar suggests that dust obscuration may be important at high redshift, and 
that current estimates for the quasar number density at $z\sim6$ are too low. 
Radio searches such as this one, and mid- to far-IR searches such as that of 
\citet{cool} are likely to find a new population of sources at high redshift.

The issue of the redshift dependence of quasar radio emission has been 
debated in the literature for decades.  The consensus ten years ago 
appeared to be that ``the fraction of radio-loud quasars decreases 
with increasing redshift'' (Schmidt et al. 1995 and references therein). 
From observations of forty optically selected quasars with $3.3<z<4.9$
(including all quasars then known with $z>3.9$) Schmidt et al. detected three 
objects at a 20~cm flux density threshold of 0.2~mJy, far below their
expectation of 9-18 detections. However, all such studies relied on rather 
small and heterogenous samples of objects with different selection criteria. 
Recently, we have studied the radio properties of the 41,295 quasars from the 
SDSS DR3 catalog that fall within the FIRST survey area 
(White et al. 2006). An important feature of our analysis is that we 
include {\it all} quasars -- especially the $\sim 90\%$ which fall below the 
radio detection threshold -- by stacking the radio images.

\citet{white06} find a very tight correlation of radio luminosity with absolute 
magnitude at all redshifts, but, importantly, the slope of the relation is not 
unity: $L_R \sim L_{opt}^{0.72}$. Thus, the higher luminosity optical objects 
-- inevitably those found in magnitude-limited high-$z$ surveys -- are
underluminous in the radio when compared with the mean of the lower-redshift 
population.  When normalized for this dependence on optical luminosity, we 
find a remarkably flat distribution of the radio-to-optical flux ratio $R$ 
for quasars with redshift: the mean value of $R$ changes by less than 0.1 
from $z=0$ to $z=5$ (White et al. 2006, figure 11; see also Cirasuolo et al. 
2006 and \citealt{petric}). The cumulative fraction of quasars above the 
Schmidt et al. (1995) limit of 0.2~mJy at 6~cm is $\sim 0.12$ in our SDSS 
sample (assuming a radio spectral slope of $\alpha = -0.5$ between 20 and 
6~cm).  Correcting for the (high) mean optical luminosity of the Schmidt et al. 
sample ($M_B = -26.3$) reduces the expected fraction by 20\% to 0.10.  Thus, 
we would expect four detections in their sample of forty; three objects were 
detected. We conclude that the expectation of discovering a radio-loud quasar 
at high $z$ is not reduced significantly by a decline in radio emission with 
redshift; indeed, the fact that two of nineteen quasars now known above 
$z=5.7$ lie above the FIRST survey threshold is not unexpected.

Our discovery of a $z>6$ radio-loud quasar in a 4-deg$^2$ survey suggests that
tractable wider-area surveys with deep $K$-magnitude (and deeper radio) limits
would be highly productive. We have not yet exhausted the possibilities in the 
FLAMEX region, as the $\sim200$ radio sources showing extended emission have 
yet to be matched.  The detection of more, higher-$z$ objects could 
provide useful probes of the epoch of reionization through redshifted
21~cm absorption measurements.

\section{Acknowledgements}

RHB and DJH acknowledge the support of the National Science Foundation under 
grants AST-05-07663 and AST-05-07598, respectively. RHB also acknowledges 
support from the Institute of Geophysics and Planetary Physics (operated under 
the auspices of the US Department of Energy by Lawrence Livermore National 
Laboratory under contract W-7045-Eng-48). This work made use of images 
and/or data products provided by the NOAO Deep Wide-Field Survey (Jannuzi and 
Dey 1999), which is supported by the National Optical Astronomy Observatory 
(NOAO). NOAO is operated by AURA, Inc., under a cooperative agreement with 
the National Science Foundation.

\clearpage

\begin{deluxetable}{lrrrc}
 \tablecaption
 {
  Photometric Observations of FIRST J1427385+331241\label{photometry}
 }
 \tablewidth{0pt}
 \tablehead{
  \colhead{Survey} & \colhead{Band} & \colhead{mag} & \colhead{error}
  & \colhead{flux density (mJy)}
 }
 \startdata
   NDWFS &  $B_w$ &   $>$26.4 & \nodata &         \\
         &  $R$   &   $>$25.5 & \nodata &         \\
         &  $I$   &     22.09 & 0.02    &         \\
  FLAMEX &  $J$   &     19.68 & 0.05    &         \\
         &  $K_s$ &     17.88 & 0.16    &         \\
     MDM &  $J$   &     19.93 & 0.50    &         \\
	     &  $K_s$ &     17.92 & 0.24    &         \\
	IRAC &  $3.6\mu$m & 16.53 & 0.11    &  0.068  \\
	     &  $4.5\mu$m & 15.91 & 0.10    &  0.077  \\
	     &  $5.8\mu$m & 15.24 & 0.09    &  0.093  \\
	     &  $8.0\mu$m & 14.86 & 0.11    &  0.072  \\
    MIPS &  $24\mu$m  &       &         & $<$0.20 \\
   FIRST &  20~cm     &       &         &  1.73   \\
 \enddata
\tablecomments{
 All magnitudes are in the Vega system.  NDWFS, MDM, and IRAC magnitudes
 were all obtained from SExtractor MAG\_AUTO. FLAMEX magnitudes are not taken 
 from the catalog, but from PSF fitting using IRAF DAOPHOT 
 (see Section~\ref{observations}).
 }
\end{deluxetable}

\clearpage

\clearpage

\begin{figure}
 \begin{tabular}{ccc}
  \includegraphics{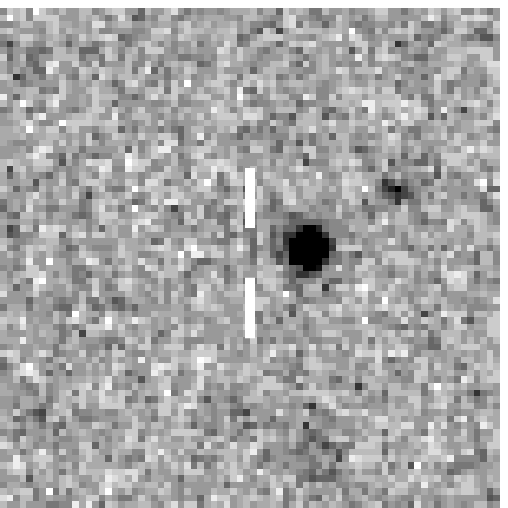}
  \includegraphics{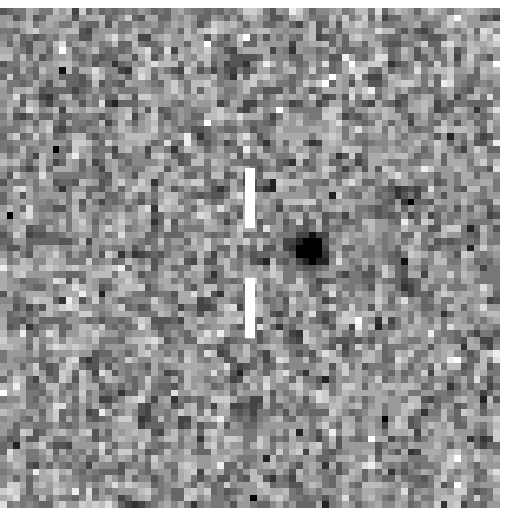}
  \includegraphics{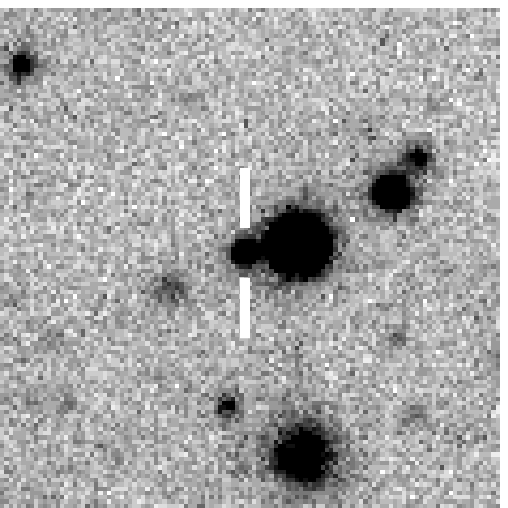}
 \end{tabular}
 \caption
{
 $30\arcsec$ SDSS $i'$, $z'$, and NDWFS $I$ cutout images of 
 images of FIRST J1427+3312. The FIRST position is marked with
 vertical lines. A $3\sigma$ detection limit of $z'=20.8$ is derived from 
 examination of the SDSS images.
}
\label{opticalcuts}
\end{figure}

\begin{figure}
 \begin{tabular}{cc}
  \includegraphics{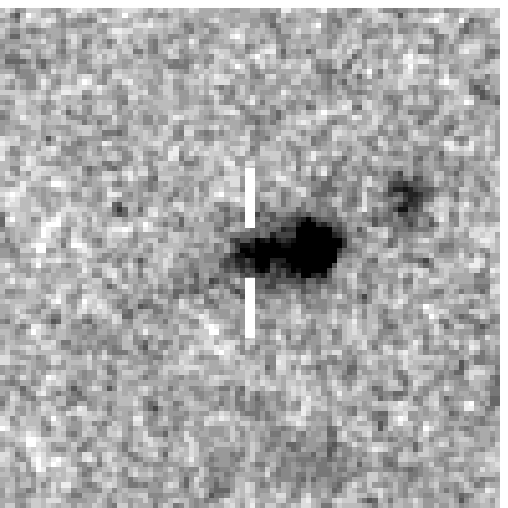}
  \includegraphics{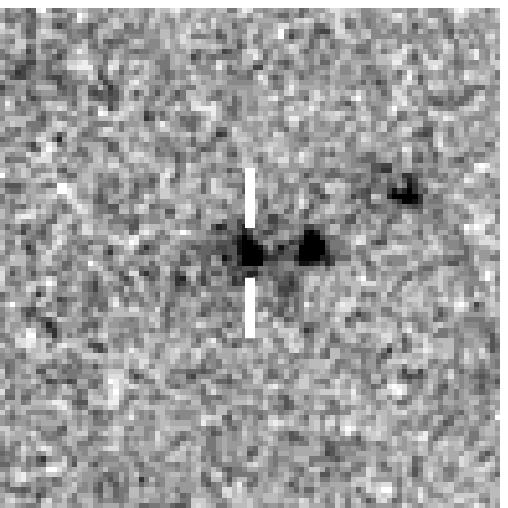}
 \end{tabular}
 \caption
{
 $30\arcsec$ FLAMEX $J$ and $K_s$ cutout images of FIRST J1427385+331241,
 with the FIRST position marked by vertical lines. The source $3\arcsec$
 west has $J=18.3$ and $K_s=18.2$ in the FLAMEX catalog, while the quasar has
 $J=18.7$ and $K_s=17.4$. Using PSF-fitting to simultaneously fit both sources
 yielded $J=19.7$ and $K_s=17.9$ for the quasar.
}
 \label{ircuts}
\end{figure}

\end{document}